# ACCELERATOR TESTS OF CRYSTAL UNDULATORS

V.M. Biryukov[@,1], A.G. Afonin[1], V.T. Baranov[1], S. Baricordi[3], S. Bellucci[2], G.I. Britvich[1], V.N. Chepegin[1], Yu.A. Chesnokov[1], C. Balasubramanian[2], G. Giannini[2], V. Guidi[3], Yu.M. Ivanov[4], V.I. Kotov[1], A. Kushnirenko[1], V.A. Maisheev[1], C. Malagu[3], G. Martinelli[3], E. Milan[3], A.A. Petrunin[4], V.A. Pikalov[1], V.V. Skorobogatov[4], M. Stefancich[3], V.I. Terekhov[1], F. Tombolini[2], U.I. Uggerhoj[5]

[1]*Institute for High Energy Physics, 142281 Protvino, Russia;* [2]*INFN - Laboratori Nazionali di Frascati, P.O. Box 13, 00044 Frascati, Italy;* [3]*Department of Physics and INFN, Via Paradiso 12, I-44100 Ferrara, Italy;* [4] *St. Petersburg Institute for Nuclear Physics, Russia;* [5]*Aarhus University, Denmark;*

**Abstract:** A series of Silicon crystal undulator samples were produced based on the approach presented in PRL **90** (2003) 034801, with the periods of undulation from 0.1 mm to 1 mm, and the number of periods on the order of 10. The samples were characterized by X-rays, revealing the sine-like shape of the crystal lattice in the bulk. Next step in the characterization has been the channeling tests done with 70 GeV protons, where good channeling properties of the undulated Silicon lattice have been observed. The photon radiation tests of crystal undulators with high energy positrons are in progress on several locations: IHEP Protvino, LNF Frascati, and CERN SPS. The progress in the experimental activities and the predictions from detailed simulations are reported.

## 1. INTRODUCTION

Many areas of modern science and engineering require intense sources of high-energy X-ray and gamma beams. Efficient sources of such radiation are electron or positron beams traversing an undulated magnetic field in accelerators. However, the usual electromagnetic undulators have a relatively large period of magnetic structure, which limits the energy of

---

generated radiation. The energy of a photon emitted in an undulator is in proportion to the square of the particle Lorentz factor γ and in inverse proportion to the undulator period L: $\hbar\omega \approx 2\pi\hbar\gamma^2 c/L$. Typically, at the modern accelerators the period of undulator in the synchrotron light sources is a few centimeters[1]. With a strong worldwide attention to novel sources of radiation, there has been broad interest [2-11] to compact undulators based on channeling in crystals, where a strong periodic electric field exists by nature. The crystalline undulators with periodically deformed crystallographic planes offer electromagnetic fields in the order of 1000 Tesla and could provide a period L in sub-millimeter range. This way, a hundred-fold gain in the energy of emitted photons would be reached, as compared to a usual undulator.

The use of bent crystals for channeling extraction of beams from accelerators has been under development at several laboratories[12-17]. A collaboration of researchers working at the 70-GeV accelerator of IHEP has achieved a substantial progress in the efficiency of crystal-assisted beam deflection: extraction efficiency larger than 85% is obtained at intensity as high as $10^{12}$ protons[17]. Since 1999, the use of bent crystals to extract beams for the high energy physics program became regular in all accelerator runs of U-70. About one half of all extracted beams with intensities up to $10^{12}$ at IHEP are now obtained with channeling crystals. The efficient practical use of crystal channeling elements for steering of particle beams at modern accelerators has been another major motivation for research towards a crystal channeling undulator.

## 2.     THE REALIZED TECHNIQUE OF CU

The first idea on how to make an undulated crystal lattice was ultrasound, suggested in 1979 by Kaplin et al[4]. Later on, in 1984, Ikezi et al[6] proposed a $Si_{1-x}Ge_x/Si$ graded composition lattice with periodical modulation of *Ge* content *x*. More recently, Avakian et al[11] suggested to apply a periodic surface strain on crystal wafer. These ideas have been theoretically developed[4-11] but they are still pending realization. Recently[2,3] we demonstrated experimentally by means of X-rays that microscratches on the crystal surface make sufficient stresses for creation of a crystalline undulator (CU) by making a series of scratches with a period of sub-millimeter range. This method is based on an interesting observation in our earlier 70-GeV proton channeling experiments ref.[12], p.120, where it was found that accidental micro-scratches on a crystal surface cause a significant deflection of high-energy particles near the scratch (see photo, Fig.1).

On this photo, a fragment of the end face of crystal and an image of the beam deflected by this crystal are presented. One can see that the image of the deflected beam strongly reflects the character of the surface scratches. This effect is explained by the fact that the protons near a scratch are channeled by deformed crystal planes. Reconstruction analysis of the angles of deflected particles shows that deformation of the crystallographic planes penetrates to substantial depths, down to a few hundred microns as depicted in Fig. 2(a). Therefore, this effect could be profitably used for creation of a CU by making a periodic series of micro grooves on the crystal surface as shown on Fig. 2(b).

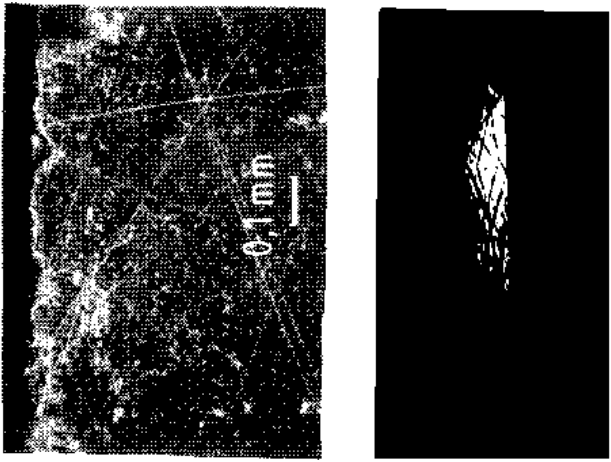

*Figure 1.* The microphotography of crystal end face with scratches (left) and image of the 70 GeV proton beam deflected by this crystal shown at a distance of 1 m (right).

Presently a series of undulators (Silicon (110) oriented wafers) was manufactured with the following parameters: the length along the beam from 1 to 5 mm, thickness across the beam 0.3 to 0.5 mm, 10 periods of oscillation with the step from 0.1 to 0.5 mm, and the amplitude on the order of 20 - 150 Å. The scratching was done in IHEP and Ferrara.

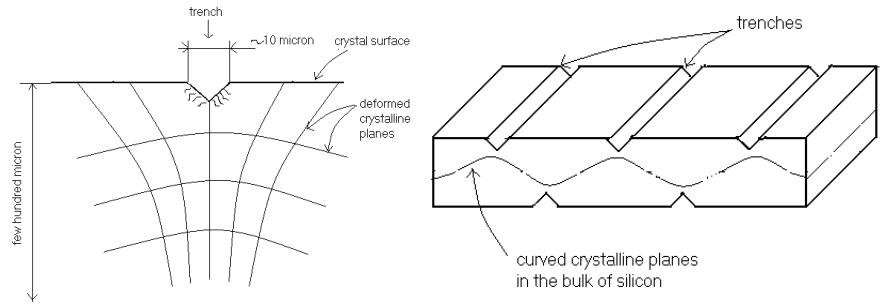

*Figure 2*. Left: The angular distortion of crystal planes near a surface scratch (groove). Right: The scheme of proposed crystalline undulator.

## 3. HIGH-ENERGY CHANNELING TESTS OF CU

Firstly, the undulators were tested and characterized with X-rays as described in refs.[2,3]. The X-ray tests showed that a sinusoidal-like shape of crystalline planes goes through the bulk of the crystal with appropriate amplitudes. In the proposed method of CU manufacturing, some part of the crystal lattice is disrupted by scratches and is not suitable for channeling, therefore we tested our crystal undulators directly for channeling of 70 GeV protons. The tests were made with our usual technique, carefully described in ref.[17]. Four CU wafers of the thickness of 0.3-0.5 mm, with about 10 periods, were bent by means of the devices described in ref.[17] by the angle of about 1 mrad and installed in a circulating accelerated beam as shown in Fig.3.

This angle of bending is sufficient to separate the circulating and the deflected (by the crystal) beams in space. The beam deflection effect due to channeling was measured by secondary emission detector, located in the vacuum chamber of the accelerator near to circulating beam.

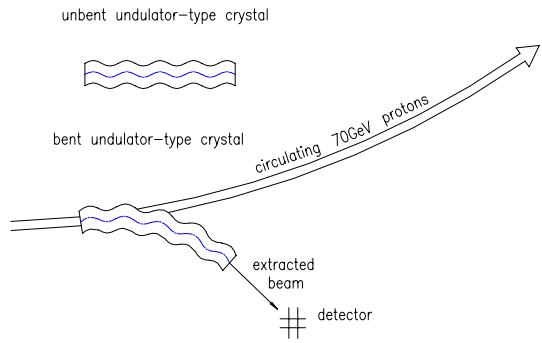

*Figure 3.* The scheme of extraction of a circulating high-energy proton beam with a bent undulator-type crystal.

On Fig. 4(a) the profile of the beam deflected by one of the tested crystals is shown as a function of the orientation of this crystal. The crystal was 5 mm along the beam and 0.3 mm across and had 10 periods of sinusoidal deformations with a step of 0.5 mm and the amplitude of about 100 Å. The top left profile corresponds to the disoriented crystal when only a scattering tail of the beam is seen. On the following picture (top right) the crystal orientation is optimal. In this case the efficiency of extraction (the ratio of the fully deflected channeled beam to the total losses of the circulating beam) is equal to 31 %. In the same conditions, a usual bent crystal (without scratches) of a similar size has extracted about 45 % of the beam. Thus the measurements showed that all CU crystals deflect protons with good efficiency and at least 70 % of the crystal cross-section is available for channeling despite of the distortions caused by scratches. Actually, part of the overall reduction of 30% should be due to additional centrifugal forces from undulation in crystal lattice, therefore the crystal cross-section available for channeling is significantly larger than 70% of the total.

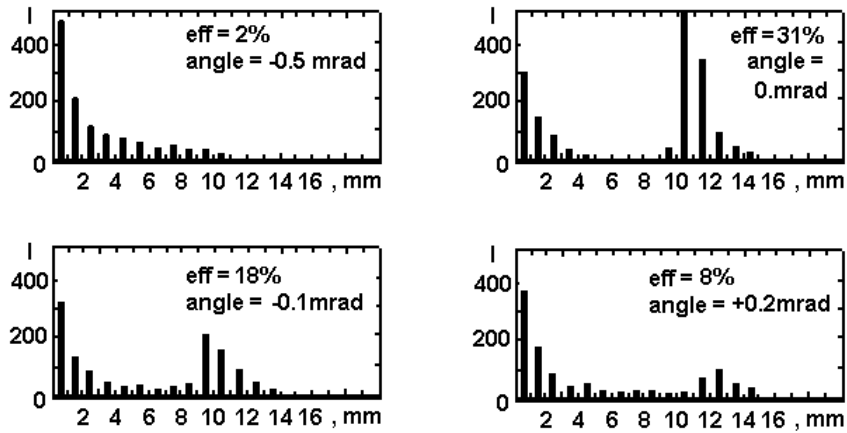

*Figure 4*. Deflected beam profile and the efficiency of extraction versus the crystal orientation angle.

The presence of periodic deformations with large amplitude in the crystal lattice, confirmed earlier by X-ray tests, and the experimentally established transparence of the lattice for channeled high-energy particles allows one to start a direct experiment on photon production from positron beam in a crystalline undulator.

## 4. PHOTON EMISSION EXPERIMENT

The present collaboration has several appropriate sites for an experiment on generation of photons in a crystalline undulator. One site is Laboratori Nazionali di Frascati with the positron beam energy 500-800 MeV. Another site is IHEP where one can arrange positron beams with the energy higher than 2 GeV. The new opportunity has occurred recently at CERN SPS positron beam line H4 under the proposal [18] at the energy of about 10 GeV and higher. The progress in LNF is described in ref. [3].

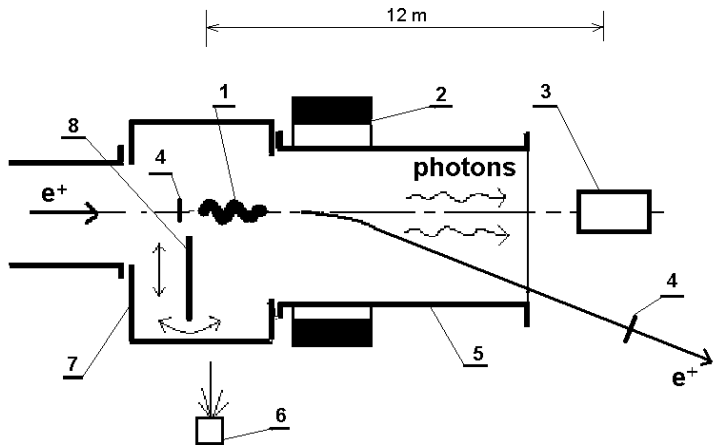

*Figure 5.* Layout of photon emission experiment at beam line 22: 1 – crystal undulator, 2 – cleaning magnet, 3 – photon detector, 4 – scintillator counters, 5 – vacuum pipe with thin exit window, 6 – laser for crystal alignment, 7 – vacuum box with flanges and electric plugs, 8 – goniometer for crystal rotation and transmission.

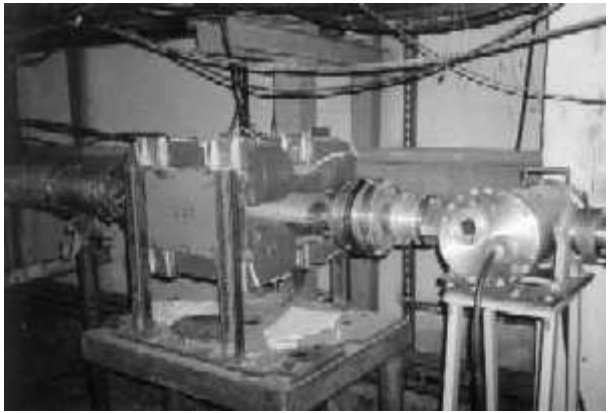

*Figure 6.* The vacuum box with CU and cleaning magnet of the setup at IHEP.

Fig. 5 shows the equipment layout for realization of experiment on the secondary beam line 22 in IHEP Protvino. Crystalline undulator is placed into vacuum box, inside which a remotely controlled goniometer is positioned. The goniometer provides a horizontal translation of the crystal for its exposure into the beam within the limits of about 100 mm, with a step size of 0.05 mm. Also, it provides an alignment of the crystal within ±30 mrad with a step size of about 0.010 mrad.

Right after the vacuum box, a cleaning magnet is positioned. The vacuum system is ended by the tube 10 m long, 200 mm in diameter, which has a

Mylar window at the end, 0.1-0.2 mm thick. Small finger scintillator counters, one installed before the crystal and another one 12 meters downstream, select beam particles in a narrow angular interval of about ±0.1 mrad and form a trigger signal for specrometer of photons. As a detector of photons, a crystal of NaI (Tl) of ∅1 cm ×10 cm is used. Part of the above described equipment is seen on photos of Figs.6, 7.

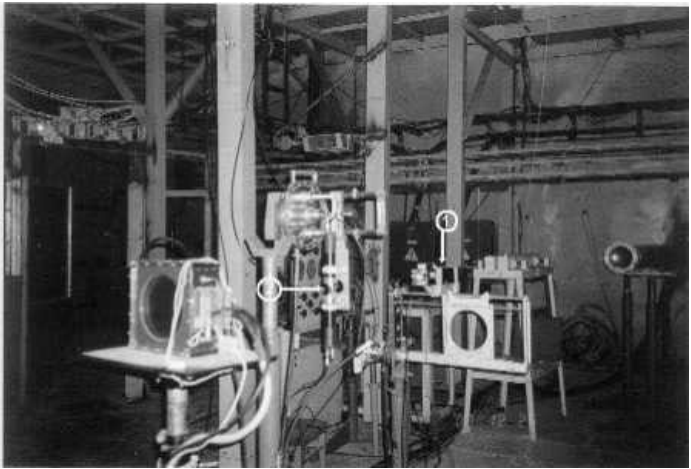

*Figure 7.* The photon spectrometer (1) and scintillator trigger (2) mounted on mechanical drivers in the setup at IHEP

The comissioning of the set-up with 3 GeV positron beam has shown that the main systems work normally and background conditions on photon detector are appropriate. The first results with necessary statistics are expected in the next run of the IHEP accelerator (planned near the end of 2004). The expected spectra of the photons generated in CU are shown in Fig. 8. Calculations were carried out for the (110) plane of a silicon single crystal with realistic sizes and take into account the following factors: channeling radiation and dechanneling process, radiation of the above-barrier positrons, absorption of gamma-quanta in the undulator bulk [3]. The calculated number of photons in the range 100–600 KeV is 0.15 per one positron channeled through the crystalline undulator. The spectral density of undulator radiation is 5 times greater than that of usual channeling radiation. Notice that our prediction (Fig.8, see also ref.[3]) is quite different from that of Solov'ev et al (e.g. Fig. 2 in ref.[9]) who predict the spectrum to be essentially zero everywhere but a very narrow undulator peak and a peak of channeling radiation, in vast contrast with our results.

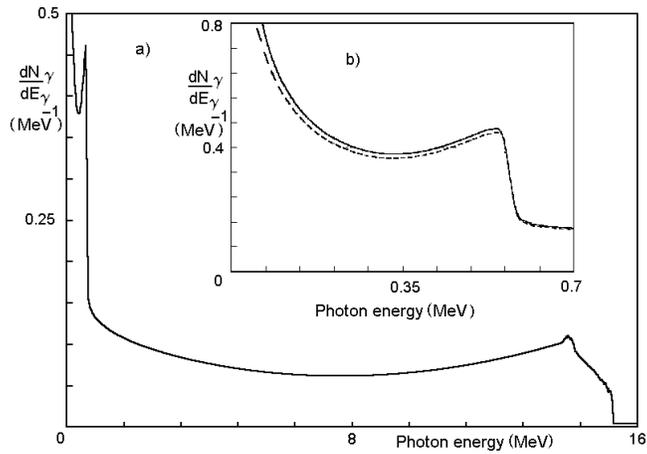

*Figure 8.* The expected photon spectrum for 3 GeV positrons in the range 0–16 MeV (a) and 0–0.7 Mev (b). The dashed curve is the photon spectrum, where absorption process of photons in the body of undulator is taken into account. The curves are normalized on one positron passing through the undulator within channeling angle.

## 5. CONCLUSION

Real crystal undulator devices have been created. The tests of these devices with X-rays and high-energy channeled protons were performed, and the experimental conditions necessary for observation of photon emission from positron beam in crystalline undulator were prepared. Creation of a new source of radiation with high spectral density, superior to the usual channeling radiation is expected on the basis of crystal undulator. Crystalline undulator would allow to generate photons with the energy on the order of 1 MeV at the synchrotron light sources where one has at the moment only 10 KeV, and for this reason crystal undulators have interesting prospects for application.

## ACKNOWLEDGEMENTS

This work was partially supported by INFN - Gruppo V, as NANO experiment, by the Italian Research Ministry MIUR, National Interest Program, under grant COFIN 2002022534, by INTAS-CERN grants 132-2000 and 03-52-6155 and by RFBR Grant No. 01-02-16229. The invitation of the talk and support from the Organizers of the workshop for the corresponding author (V.M.B.) is gratefully acknowledged. Armenian hospitality has greatly contributed to the success of the workshop.